# A Link Quality Model for Generalised Frequency Division Multiplexing

Ghaith R. Al-Juboori[(1)], David Halls[(2)], Angela Doufexi[(1)] and Andrew R. Nix[(1)]

[(1)]Communication Systems & Networks Group, Department of Electrical and Electronic Engineering

University of Bristol, Bristol, United Kingdom

[(2)]Telecommunication Research Laboratory, Toshiba, Bristol, United Kingdom

Email: {Ghaith.Al-juboori, Angela.Doufexi, Andy.nix}@bristol.ac.uk, david.halls@toshiba-trel.com

*Abstract* − 5G systems aim to achieve extremely high data rates, low end-to-end latency and ultra-low power consumption. Recently, there has been considerable interest in the design of 5G physical layer waveforms. One important candidate is Generalised Frequency Division Multiplexing (GFDM). In order to evaluate its performance and features, system-level studies should be undertaken in a range of scenarios. These studies, however, require highly complex computations if they are performed using bit-level simulators. In this paper, the Mutual Information (MI) based link quality model (PHY abstraction), which has been regularly used to implement system-level studies for Orthogonal Frequency Division Multiplexing (OFDM), is applied to GFDM. The performance of the GFDM waveform using this model and the bit-level simulation performance is measured using different channel types. Moreover, a system-level study for a GFDM based LTE-A system in a realistic scenario, using both a bit-level simulator and this abstraction model, has been studied and compared. The results reveal the accuracy of this model using realistic channel data. Based on these results, the PHY abstraction technique can be applied to evaluate the performance of GFDM based systems in an effective manner with low complexity. The maximum difference in the Packet Error Rate (PER) and throughput results in the abstraction case compared to bit-level simulation does not exceed 4% whilst offering a simulation time saving reduction of around 62,000 times.

## I. Introduction

The requirements for 5G systems vary depending on the scenario considered, such as Internet of Thinks (IoT), Machine Type Communication (MTC) and high data rate mobile communications. Different techniques need to be deployed to achieve these requirements, including Massive MIMO, millimetre wave bands and using new physical layer waveforms. The selection of the air interface is key due to its impacts on the transceiver complexity and the system level performance [1].

Orthogonal Frequency Division Multiplexing (OFDM) is successfully used in many wireless standards such as Wireless Local Area Networks (WLANs) and the 4G cellular mobile standards (LTE & LTE-A). This success is due to its desirable features such as robustness to Inter-Symbol Interference (ISI) and low implementation complexity due to the efficient use of Inverse Fast Fourier Transform/Fast Fourier Transform (IFFT/FFT) processing [2]. On the other hand, OFDM suffers from several disadvantages for example, its high out of band radiation, high sensitivity to Carrier Frequency Offset (CFO) and high peak to average power ratio (PAPR) [3]. These drawbacks may prevent it from being used in 5G systems.

Recent research looking into the selection of a new air interface for 5G has focused mainly in two areas. The first has proposed enhancements and alternatives to the OFDM waveform, in order to improve many of its features such as the spectral containment and the sensitivity to CFO, as in [4]. The second area is looking at alternative waveforms to OFDM. Many candidates have been proposed such as Generalised Frequency Division Multiplexing (GFDM), Filter Bank Multi-Carrier (FBMC) [5] and Universal Filtered Multi-Carrier (UFMC) [6]. In this paper, we focus on the GFDM waveform. System-level performance studies are necessary to accurately evaluate the performance of a system using a GFDM waveform, however, these studies have high computational complexity if they are implemented using bit-level simulators.

In this paper, the link quality model, which is often used to evaluate the system-level performance for OFDM [7], in a simple and low complexity manner, is investigated for GFDM. To the best of our knowledge, this subject is not investigated yet. The Mutual Information (MI) based link quality model is used because it outperforms other models, as illustrated in section III. The remainder of this paper is organised as follows: in section II, a brief description of the GFDM air interface, its features and a low complexity transceiver model are given. In section III, general descriptions of the link quality model and the MI-based link quality model (for OFDM & GFDM) are presented. The simulation parameters which have been used in this paper are listed in section IV. The results are shown and discussed in section V. Finally; the conclusions are given in section VI.

## II. GFDM System Model

### A. GFDM Overview

GFDM is a digital multicarrier modulation scheme, and its flexibility helps it to address the different requirements of 5G; the basic structure for the GFDM transmitter is illustrated in Fig. 1. The GFDM block consists of *K* subcarriers and *M* sub-symbols per subcarrier, unlike OFDM which has only one symbol per subcarrier. A pulse shape filtering process is used on each subcarrier to reduce the Out-Of-Band (OOB) radiation.

Different types of filters (orthogonal and non-orthogonal) can be used as a prototype filter, and this increases the flexibility of the GFDM waveform [8]. An up-conversion process is performed before adding the sub-carriers signals together to form the final GFDM signal. The GFDM signal can be expressed as:

$$x[n] = \sum_{k=0}^{K-1} \sum_{m=0}^{M-1} g_{k,m}[n] d_{k,m}, \quad (1)$$

where $d_{k,m}$ is the complex data symbol which is transmitted on the sub-carrier $k$ and the sub-symbol $m$. $g_{k,m}$ represents the time and frequency shifted version of the impulse response of the prototype filter, it can be written as:

$$g_{k,m}[n] = g[(n-mK) \bmod N] e^{-j2\pi \frac{k}{K} n}, \quad (2)$$

where $n$ is the sampling index ($n=0,......, N$-1) and $N$ is equal to $K$ by $M$.

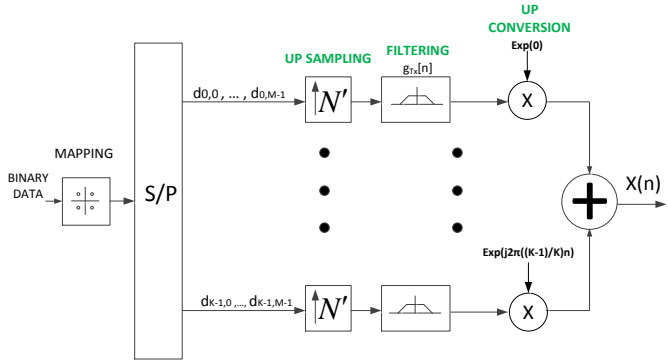

Fig. 1: The Basic GFDM Transmitter [9].

Direct implementation of the two above equations (1 and 2) requires a number of complex multiplications which is equal to $NKM^2$. However, in this paper, the method used in [3] and implied a significant reduction in the computation complexity by reformulating the GFDM transmitter in a similar fashion to that used in OFDM (employing an IFFT/FFT), is applied.

After the GFDM modulator, the cyclic prefix ($N_{CP}$ samples) is added to the GFDM signal. One of the important reasons for this addition is in order to be able to perform the equalisation process at the receiver side in the frequency domain. After that, the signal is transmitted through the channel. Assuming perfect synchronisation and channel estimation processes, the reverse steps are applied to get the estimated data sequence at the receiver. Several methods can be used to implement the GFDM demodulator such as a matched filter, zero forcing and Minimum Mean Square Error (MMSE); for more details, please see [1]. The zero forcing method is applied in the paper.

### III. LINK QUALITY MODEL

Recently, the link quality model (a PHY abstraction method) has been effectively employed for evaluating system performance and in predicting link adaptation precisely based on the Signal to Interference and Noise Ratio (SINR) measured by the receiver [10]. The link quality model comprises a vector of received SINR, the post processing SINR across the coded block at the input of the decoder for certain channel realisation is mapped into a single value which is called the Effective SINR (ESINR). Using this value, the model can predict the Block Error Rate (BLER), i.e, the Packet Error Rate (PER), for a given channel snapshot across the OFDM subcarriers which are used to transmit the coded block. Fig. 2 illustrates the basic concepts of the abstraction approach which is explained in detail in the following. Firstly, the post processing SINR per sub-carrier $n$ (frequency sample) for a certain user $i$ is calculated as [11]:

$$SINR^i(n) = \frac{P_{tx}^{(i)} \cdot P_{loss}^{(i)} \cdot |H^{(i)}(n)|^2}{\sigma^2 + \sum_{\substack{q=1 \\ q \neq i}}^{N_I} P_{tx}^{(q)} \cdot P_{loss}^{(q)} \cdot |H^{(q)}(n)|^2}, \quad (3)$$

where $q$ represents the interferer, $N_I$ is the total number of interferers, $P_{tx}$ is the transmitted power and $P_{loss}$ is the path loss including shadowing. Secondly, the abstraction transforms the vector of SINR for a certain block (OFDM block) using a mapping function $\Phi$ (SINR) to another domain, which is related to the mapping function. After that, the transformed values are linearly averaged over the block before the average value is returned back to the SINR domain to get the ESINR ($\gamma_{eff}$) using $\Phi^{-1}$ as shown in the following equation:

$$\gamma_{eff} = \Phi^{-1} \left[ \frac{1}{J} \sum_{j=1}^{J} \Phi(\gamma_j) \right], \quad (4)$$

where $J$ is the number of sub-carriers and $\gamma_j$ is the SINR for the sub-carrier $i$. Different methods for mapping have been discussed in the literature, namely the Exponential Effective SINR mapping, where $\Phi$ is replaced by the negative exponential function and Mutual Information Effective SINR Mapping (MIESM). In this paper, the MIESM method is proposed due to its simple structure and high accuracy compared to the other methods [10]. The details of this approach will be given in the next sub-section. Finally, the ESINR ($\gamma_{eff}$) will be used to calculate the BLER based SNR versus BLER curves in the Additive White Gaussian Noise (AWGN) case. This technique has been widely validated for OFDM waveform in different works, for example [10, 11].

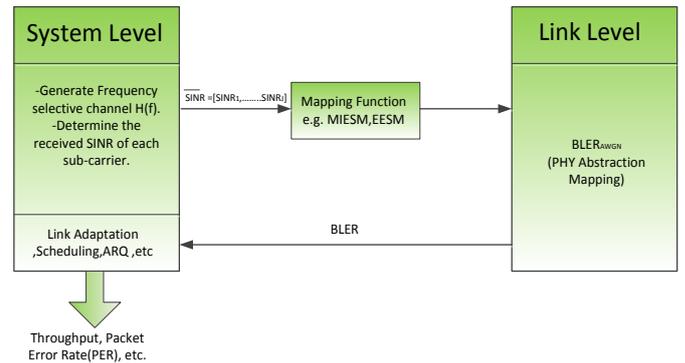

Fig. 2: PHY link-to-system mapping procedure.

### A. Mutual Information Based Link Quality Model

As mentioned in [10], the MI-based PHY abstraction technique can be separated into modulation and coding models. Accordingly, Fig. 3 shows the MI-based link quality model structure, and a brief description for each model is given below:

1-Modulation Model

In this model, the description of the maximum channel capacity of a specific modulation scheme is given based on a symbol-by-symbol basis without considering the decoding information loss. The Symbol Information (SI) of the channel symbol, for a given SNR value (γ), is expressed as:

$$SI(\gamma, m) = E\left[log_2 \frac{P(Y|X,\gamma)}{\sum_X P(X)P(Y|X,\gamma)}\right], \quad (5)$$

where $E$ is the expected value, $Y$ is the complex value channel output symbol with SNR equal to γ, $m$ is the modulation order, $P(Y|X,\gamma)$ is the AWGN channel transition probability density conditioned on the noiseless channel symbol $X$, and $P(X) = \frac{1}{2^m}$ is assumed.

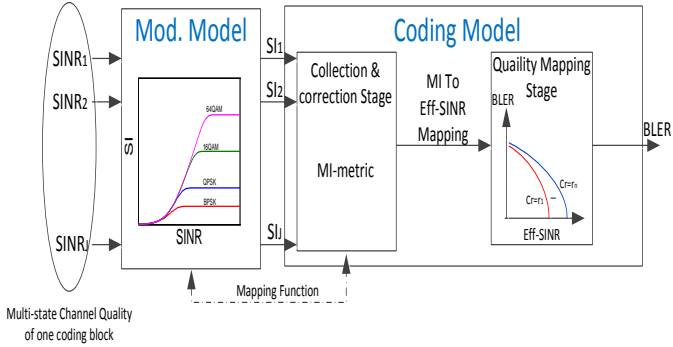

Fig. 3: MI-based quality model structure.

2-Coding Model.

This model contains two stages; the SI collection /correction stage and the quality mapping stage. In the first stage, the SI of $J$ symbols in each block are collected/corrected and added together to get the Received coded Bit Information (RBI). These symbols have SINR values of $\{\gamma_1, \gamma_2, ..., \gamma_J\}$ and modulation orders of $\{m_1, m_2, ..., m_J\}$ and the RBI's calculation is expressed as:

$$RBI = \sum_{j=1}^{J} SI(\gamma_j, m_j) \quad (6)$$

The Received Bit Information Rate (RBIR), which is equivalent to the sample average of the normalised SI over the received block for code blocks for a given modulation, with a value in the range [0,1], can be evaluated as:

$$RBIR = RBI / \sum_j^J m_j \quad (7)$$

To take the practical coding loss from the Shannon limit into consideration (the correction process in the first stage), the SI values can be multiplied by ($\gamma_{code}$) before they are combined. As stated in [10], $\gamma_{code}$ is close to 1 for Turbo and convolutional codes. Finally, the RBIR value is then mapped again to the SINR domain to get the ESINR which will be used to get the BLER value based on the AWGN look-up table. The coding model only relates to the performance of the coding system in AWGN, decoding algorithm and block size.

*B. Mutual Information Based Link Quality Model for GFDM*

According to the method which is used to implement the GFDM transceiver in this study [3], the $M$ data sub-symbols are firstly converted to the frequency domain by taking the FFT. Since this process distributes the $M$ sub-symbols on $M$ frequency samples, therefore, the SI distribution is assumed to be uniform over the frequency samples. Based on this assumption, the same steps can be used for calculating MI in the GFDM case as are used for OFDM waveforms [10].

## IV. SIMULATION PARAMETERS

*A. AWGN and Rayleigh channel models*

Here, a comparison between the simulation results and the PHY abstraction results in the case of AWGN and narrowband Rayleigh is shown. A cyclic prefix is used to prevent ISI. The parameters that are used in this case are listed in Table I.

TABLE I: SIMULATION PARAMETERS

| Parameter | Value |
|---|---|
| No. of sub-carriers | 64 |
| No. of sub-symbols | 9 |
| Filter types | Dirichlet, RC-0.1, RC-0.9 |
| Channel types | AWGN, Narrow band Rayleigh |
| Channel coding | Turbo code |
| MCS modes | QPSK-1/3, 16QAM-1/3 |

*B. System level parameters*

A comparison between the system-level results for LTE-A based on the GFDM waveform using the bit-level simulator, which is already done by the authors as a part of a previous study [9], and the PHY abstraction method is also presented. A 3GPP macro-cellular deployment with a frequency reuse factor of one is used. There are three sectors in each cell, and the cell radius, cell diameter and Inter-Site Distance (ISD) are $R$, $2R$ and $3R$ respectively [12]. The User Equipment (UEs) locations were randomly distributed at the street level in the cell and at a distance between 50-1000 m from the main Base Station (BS). The 3D extended 3GPP-ITU channel model has been used, where the effect of the elevation is also taken into consideration [13]. Table II summarises the system level parameters that have been used in this case. One thousand channel snapshots have been produced for each link (between each UE and the main BS and each UE and each one of the other six first-tier interfering BSs) to get statistically relevant performance results. The GFDM parameters for this case are listed in Table III.

TABLE II: SYSTEM-LEVEL PARAMETERS

| Parameter | Value |
|---|---|
| Channel model | Extended 3D 3GPP-ITU (SISO) |
| PDSCH simulation models | Bit level Simulator & PHY Abstract |
| Bandwidth | 20 MHz |
| Carrier Frequency | 2.6 GHz |
| Environment | Urban-Macro |
| Cell Radius | 500 m |
| BS transmit power | 43 dBm |
| No. of users per cell | 900 |
| BS antenna height | 25 m |
| Antennas | Measured patch BS & UE handset as in[14] |
| BS down tilt | 10 ° |
| Minimum user sensitivity | -120 dBm |
| Link direction | Downlink (from BS to UE) |
| Noise Figure | 9 dB |

TABLE III: GFDM PARAMETERS

| Parameter | Value |
|---|---|
| Sub-frame duration | 1ms or 30,720 samples |
| GFDM symbol duration | 66.67μs or 2048 samples |
| Sub-symbol duration | 4.17μs or 128 samples |
| Subcarrier spacing | 240 kHz |
| Sampling frequency | 30.72 MHz |
| Total No. of sub-carrier ($K$) | 128 |
| No. of active subcarriers ($K_{on}$) | 75 |
| No. of sub-symbols per GFDM symbol ($M$) | 15 |
| No. of GFDM per sub-frame | 15 |
| Cyclic prefix length | 4.17μs or 128 samples |
| Prototype filter | Dirichlet |
| Channel coding | Turbo code |
| MCS modes | QPSK1/3, 16QAM1/2, 64QAM2/3 |

## V. RESULTS

### A. Comparison using AWGN and Rayleigh channel models.

Fig. 4 shows the BER versus SNR performance for two Modulation & Coding Schemes (MCSs) and three types of filters for the bit-level and PHY abstraction methods in an AWGN channel. As we can see, the PHY abstraction results closely match the simulation results. In this case, the channel frequency samples are equal to one (AWGN channel), and the SNR per each frequency sample will be equal (no interference between the UEs is assumed). This means that the mapping, averaging and quality mapping processes (look-up table) are working properly based on the frequency sampling. Moreover, we see a difference depending on the filter type used at each MCS. For example, the Dirichlet filter has the best performance compared to the RC filters due to the absence of Inter-Carrier Interference (ICI). Additionally, there is a degradation in the RC filter's performance due to ICI. This degradation depends on the roll-off factor of the filter, for example, the difference is fairly negligible in the case of a roll-off factor of 0.1 when compared to the orthogonal filter, whilst it becomes around 2 dB in the case of a roll-off factor of 0.9.

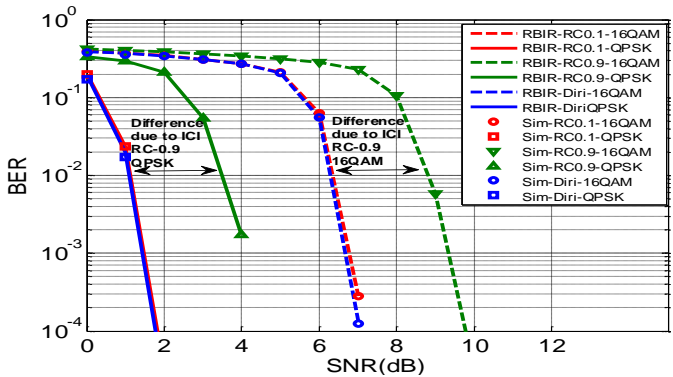

Fig. 4: Performance of the two approaches in AWGN.

Fig. 5 illustrates the performance of GFDM in a narrowband Rayleigh channel. The channel frequency response, in this case, is flat, this means that the SNR per frequency sample in each block will be equal. As can be seen, results show a very good match between the two approaches. Furthermore, a difference in performance depending on the filter type is also seen.

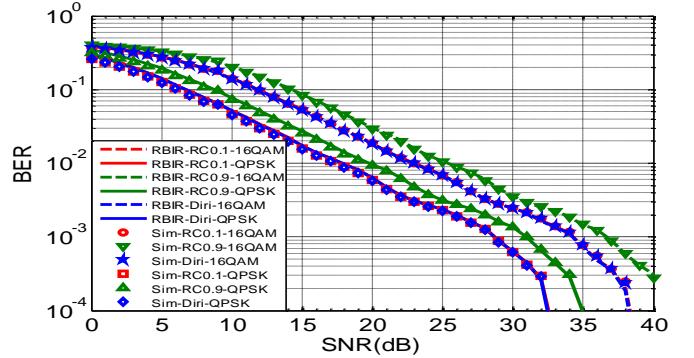

Fig. 5: Performance Comparison in narrowband Rayleigh

### B. System-level analysis

In order to represent realistic channel scenarios for system-level analysis, the 3D-3GPP ITU channel model is used here. Both PER and throughput metrics are shown in the performance evaluation since they are common metrics used in system-level studies. Fig. 6 illustrates the performance of the two approaches (the bit-level simulation and the PHY abstraction), PER versus the SNR at a certain UE location. As mentioned in [10], the accuracy term, which is used to measure the accuracy of the PHY abstraction method, is defined as the maximum SINR difference between the simulated and the predicted results at BLERs from 1% to 10%. Table IV lists the accuracy term for different MCSs. It can be seen that the maximum difference is around 0.6 dB in the 64QAM-2/3 MCS. The above results are for a unity adjusting factor ($\gamma_{code}$). However, we found that in this case, the best value of $\gamma_{code}$ depends on the channel type, i.e. it is not the same value for different channel types.

TABLE IV: Accuracy for different MCSs

| MCSs | Accuracy 1~10% BLER |
|---|---|
| QPSK-1/3 | 0.5 dB |
| 16QAM-1/2 | 0.5 dB |
| 64QAM-2/3 | 0.62 dB |

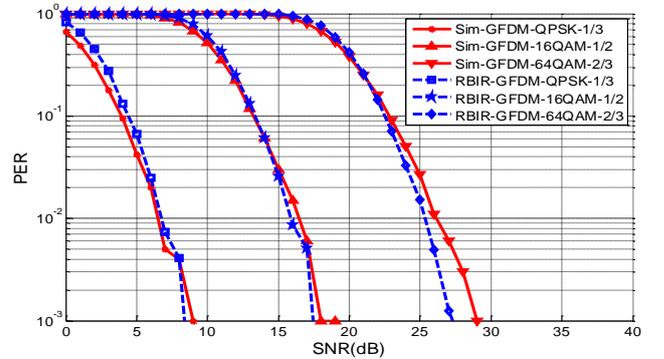

Fig. 6: Performance Comparison for certain UE.

Fig. 7 represents the Cumulative Distribution Function (CDF) of the PER for the UEs in interference-free (SNR) and interference present (SINR) cases in both methods. The effect of taking the interference into consideration which leads to increase PER is clearly seen in this figure. Moreover, the results of the two methods are close, and the maximum difference between them is around 4%.

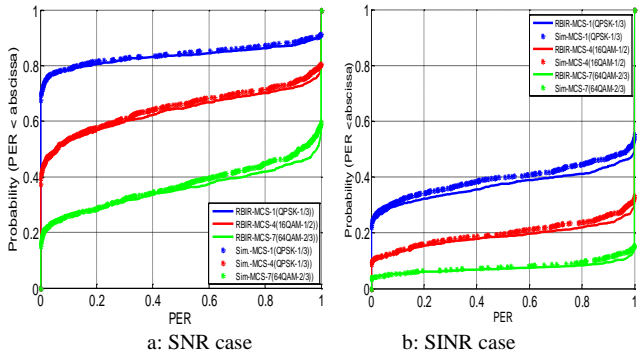

a: SNR case    b: SINR case
Fig. 7: CDF of UEs PER in SNR and SINR cases

Fig. 8 shows the CDF for the throughput for both approaches in interference-free and interference-included cases; given the use of adaptive MCS selection, i.e. for each user the best MCS mode is selected. It can be observed that the simulation and PHY abstraction results are very similar. The throughput for both approaches is clearly much better in the interference-free case. It can be seen that 65% of the UEs have a throughput greater than 20 Mbps in the interference-free case; while just 20% of the UEs achieve this rate when interference is considered in the simulator. However, the difference between the two methods is less in the interference-included case, which is the more realistic case. Additionally, the maximum difference in the throughput's CDF values in both cases, interference-free and included, does not exceed 4%.

Finally, the total time required to run the system-level simulation using the PHY abstraction on a PC was 1.59 hours. The expected time required to run the full bit-level simulation on a PC-based is around 98,000 hours (although the simulation was actually executed on the High-Performance Computing platform at the University of Bristol). This means that around 62,000 times saving in time can be obtained.

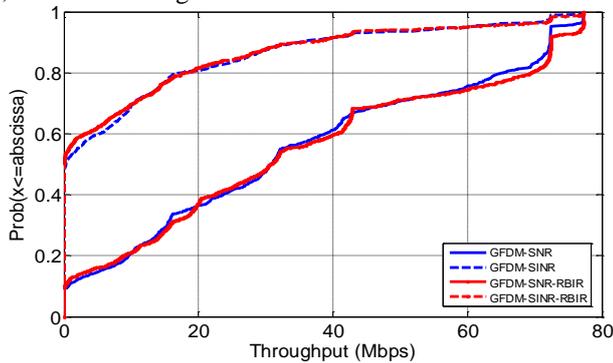

Fig. 8: CDF of UEs Throughput in SNR and SINR cases.

## VI. CONCLUSION

This paper has proposed an MI-based link quality model for the GFDM waveform. As the simulation results show, the results of the bit-level simulator and the PHY abstraction model are very closely matched. Moreover, a system-level study in a realistic channel scenario was presented for GFDM.

These results demonstrate that the MI-based link quality model (PHY abstraction) can be used effectively in the implementation of GFDM based system system-level studies and can lead to a significant reduction in the computational complexity. This will save time and resources required to measure and study GFDM performance and to analyse its suitability as a new waveform for 5G systems.

ACKNOWLEDGMENTS

Ghaith Al-Juboori would like to thank the Higher Committee for Education Development (HCED) in Iraq, Ministry of Oil and the University of Baghdad for sponsoring his PhD studies.